# Reverse depth profiling of electrodeposited Co/Cu multilayers by SNMS

A. Csik[1*], K. Vad[1], G.A. Langer[2], G.L. Katona[2], E. Tóth-Kádár[3], L. Péter[3]

[1]*Institute of Nuclear Research of the Hungarian Academy of Sciences, H-4001, Debrecen, P.O. Box 51, Hungary*

[2]*Department of Solid State Physics, University of Debrecen, H-4010, Debrecen, P.O. Box 2, Hungary*

[3]*Research Institute for Solid State Physics and Optics of the Hungarian Academy of Sciences, H-1525, Budapest, P.O. Box 49, Hungary*

**Abstract**

The overall quality of multilayer thin films prepared by electrodeposition is strongly influenced by the surface and interface roughness which increases with the layer number. For that very reason the reliable analysis of the first few layers can be necessary. However, in depth profiling methods based on sputtering techniques the first layer is always found at the bottom of the sputtered crater. Since the depth resolution decreases during sputtering, the analysis of the first few layers are difficult. In order to circumvent this problem, we used reverse Secondary Neutral Mass Spectrometry (SNMS) depth profiling method for electrodeposited multilayered films. We prepared thin film samples in two ways. First, Co/Cu multilayer stacks were electrodeposited on Si/Cr/Cu substrates and SNMS depth profiling was carried out from the direction of the topmost layer. Secondly, elecrodeposited Co/Cu multilayer stacks were coated with a few microns thick Ni layer and detached from the Si substrate in order to study the film structure from the side of the substrate. Using this latter method, we were able to analyze the first and, probably, the most even layers of the thin film structure with high resolution.

---

[*] Corresponding author. *Fax*: +36-52-416181
*E-mail address*: csik@atomki.hu (A. Csik)

**1. Introduction**

Magnetic/non-magnetic multilayers with a bilayer thickness of a few nanometers have been widely investigated since they show the giant magnetoresistance (GMR) effect [1,2]. Nanostructured magnetic multilayers can be prepared by a variety of thin film deposition techniques, such as magnetron sputtering, evaporation or molecular beam epitaxy (MBE). Among these, electrodeposition is considered to be one of the simplest techniques for the deposition of nanoscale metallic multilayers [3,4]. However, in contrast to physical vapour deposition techniques, which require high vacuum systems, electrodeposition takes place in a reactive environment, which means that the surface composition of the sample may change after deposition. The overall quality of layered structures (consisting of 3 or more different materials) or multilayers is dominantly determined by the first few layers and in most cases the quality of the interfaces changes as the number of layers increases. There are many methods to improve the quality of the layers (e.g. ion bombardment, optimization of the substrate temperature) but it is quite difficult to employ them during electrodeposition. For that reason in all cases it is very important to get reliable information about the quality of the layered structure in case of samples prepared by electrodeposition.

Depth profiling is one of the most powerful methods for the analysis of thin film multilayer structures, in particular for the determination of the depth distribution of components in thin film materials. Among depth profiling methods Secondary Neutral Mass

Spectrometry (SNMS) is a suitable technique for depth distribution measurement of the constituents of thin films [5]. Quantifiability is one of the major merits of SNMS, in contrast to other depth profiling methods, where preferential sputtering makes quantitative analysis more difficult.

In our previous work [6] we reported a successful application of SNMS for characterization of electrodeposited CoCu/Cu and CoNiCu/Cu multilayers by measuring the composition gradient along the growth direction of thin films. As we found, the depth resolution decreased significantly with the increase of crater depth. In order to clarify the origin of this significant decrease, we investigated the possible source of the low depth resolution. In SNMS depth profiling using low ion energy for sputtering, two phenomena can cause low depth resolution: bad quality of the crater profile [7], and high surface and interface roughness. The first assumption that the low depth resolution was caused by inappropriate crater profile could be easily excluded because the crater profile of all the studied samples was almost ideal. As an example, in Fig. 1 we show a crater profile measured on Sample B in Reference [6]. In spite of the good crater shape even at 300 nm depth, the depth resolution was very low (see Fig. 2 in Ref. [6]). We suppose that this is due to the concentration inhomogeneity arising from the sample preparation method and surface roughness. The aim of our work is to show a method by which the first few layers of an electrodeposited multilayer stack can be analyzed in a better way than in classical arrangement.

## 2. Sample preparation and measurements

Reverse depth profiling simply means that the layer structure of a multilayer stack is studied from the direction of the substrate. In order to examine the applicability of the reverse depth profiling to the aforementioned problem, we performed measurements on samples of two kinds. First, Co/Cu multilayer stacks were electrodeposited on Si/Cr/Cu substrates and SNMS depth profiling was carried out from the direction of the topmost layer. Secondly, electrodeposited Co/Cu multilayer stacks were coated with a few microns thick Ni layer which made it possible to remove the self-supporting deposit with the evaporated Cr/Cu layer from the Si substrate and to study the film structure from the direction of the substrate. Using this latter method, we were able to analyze the first and, probably, the most even layers of the thin film structure.

The Co/Cu multilayer was electrochemically deposited by the two-pulse plating method from the following electrolyte: 0.8 mol/liter $CoSO_4$, 0.015 mol/liter $CuSO_4$, 0.2 mol/liter $H_3BO_3$ and 0.2 mol/liter $(NH_4)_2SO_4$. Electrodeposition was performed in a tubular electrochemical cell, where the cathode was placed at the bottom in horizontal position. A detailed cross-sectional view of the electrochemical cell and the electrode arrangement is contained in Ref. [8]. The cathode was a metal-coated Si wafer in which a thin buffer layer of Cr (20 nm thick) and a seed layer of Cu (20 nm thick) were deposited at room temperature on (111) oriented silicon substrates by evaporation. While the Cr layer assured sufficient adherence to the Si wafer during the deposition, the Cu layer served as a seed layer for the first electrodeposited Cu layer. The multilayer stacks were deposited in the G/P mode [9], i.e. the magnetic Co layer was produced with a high constant current of -60 $mA/cm^2$ (galvanostatic or G mode), and the non-magnetic Cu layer was deposited at low cathodic potential (potentiostatic or P mode), at -600 mV vs. a saturated calomel reference electrode. After preparation of the required thin film structure the electrolyte was changed to a Watts type Ni plating electrolyte and the Co/Cu multilayer was coated with a 3 µm thick Ni layer whose mechanical toughness made it possible to remove the whole metallic thin film from the Si wafer. The disadvantage of this method was that the boundary between the Co/Cu multilayer and the Ni coating was a bit smeared out because of the corroding effect of the electrolyte used for the deposition of the Ni supporting layer.

After the sample preparation, the Si wafer (0.26 mm thick) was broken at the middle line and the film was simply pulled off from it. In present experiment we performed measurements on a number of samples with nominal compositions of Si/Cr(20nm)/Cu(20nm)/[Co(5.4nm)/Cu(4.4nm)]x7, but in this paper we present the results measured on two of them. One was peeled off from the substrate with the help of the thick Ni layer (sample „A"), the other multilayer sample was not coated with Ni and remained on the Si wafer (sample „B").

The depth profile analysis of the samples was performed by SNMS in Direct Bombardment Mode (type INA-X, SPECS GmbH, Berlin), as described in a previous paper [10]. In order to achieve high depth resolution, 350 eV $Ar^+$ ions were used for sputtering. The erosion area was confined to a circle of 2 mm in diameter by means of a Ta mask. The lateral homogeneity of the ion bombardment was examined by the measurement of the sputter crater with a profilometer (Ambios Technology, 1 nm depth resolution) after each run.

## 3. Results and discussion

The surfaces of the sample „A" and the Si wafer were checked by optical microscope. We found that although the detachment of the film from the Si wafer occurred at the boundary of the Si wafer and the Cr layer, the surface of the Cr film was not smooth and some Cr moieties remained on the Si crystal. Since this effect caused an extra surface roughness which would have decreased the depth resolution due to the intermixing of the emission of various layers, we performed our measurements on those areas of the sample where the surface was perfect.

Figure 2 shows the depth profiles of sample „A" and sample „B". The relative signal variation is much higher for sample „A", i.e. when sputtering was started at the Cr layer. The increased layer resolution is also accompanied with lower noise. The depth profile obtained with the reverse sputtering direction resolves the layer structure in the vicinity of the Si wafer much better than the conventional depth profile analysis for the other sample. The better quality of depth resolution due to the reverse sputtering was experienced in each sample we investigated. So, on the statistical basis one can conclude that reverse depth profiling gives better results for the layers of an electrodeposited multilayer stack. We are aware that the best comparison between the two depth profiling methods would be performed on the same sample, but unfortunately the small volume of our electrochemical cell did not provide us an opportunity to produce samples with large enough dimensions up till now. Also, the peel-off procedure of the samples works the better the smallest the sample surface area.

In our SNMS experiments the depth resolution was mainly determined by the surface roughness. The other effects of ion bombardment, e.g. knock-on-effects and atomic mixing, were significantly diminished by the low energy of bombarding $Ar^+$ ions. The surface roughness was analyzed by both surface profiler and Atomic Force Microscope (AFM). (The detailed results of AFM measurements will be published in a separate paper [11].) We found that the height distribution of surfaces could be fitted perfectly by a Gaussian-function, and that the full width at half maximum (FWHM) value of the Gaussian-function which characterizes the surface roughness was increasing during sputtering. At a hundred nm sputtering depth, the FWHM increased from the initial 2 nm value up to 8 nm which is higher than the layer thicknesses. In our experiments we experienced surface roughness dependence as a function of the sample thickness, too.

## 4. Conclusion

We demonstrated by our measurements that SNMS is a very useful method for analyzing multilayer samples produced by electrochemical deposition. Applying reverse order sputtering for depth profile analysis, the layer structure of electrochemically deposited Co/Cu multilayers can be studied with better resolution than using the conventional depth profile analysis where the sputtering is started at final sample surface. The results can bear important information on the electrochemical layer growth processes in nanometer scale especially at the beginning of the deposition. It is necessary to emphasize that this method is applicable only for layers, where the peel-off technique works.

**Acknowledgement**

This work was supported by the grant of Hungarian Scientific Research Fund (Grant number OTKA K-60821). The authors are very grateful to Gy. Molnár (Research Institute for Technical Physics and Materials Science, Hung. Acad. Sci.) for preparing metal coated substrates.

**References**

[1] Berkowitz AE, Mitchell JR, Carey M J,Young AP, Zhang S, Spada FE, Parker FT, Hutten A, Thomas G. Phys Rev Lett 1992;68:3745.

[2] Nabiyounia G, Schwarzacher W, Rolikc Z, Bakonyi I. Journal of Magn Magn Mat 2002;253:77.

[3] Ross CA. Ann Rev Mater Sci 1994;24:159.

[4] Schwarzacher W, Lashmore DS. IEEE Trans Magn 1996;32:3133.

[5] Oechsner H, Reihert L. Phys Lett 1966;23:90.

[6] Katona GL, Berényi Z, Péter L, Vad K. Vacuum 2008;82:270.

[7] Stumpe E, Oechner H, Schoof H. Appl. Phys. 1979;20:55

[8] Péter L, Pádár J, Tóth-Kádár E, Cziráki A, Sóki P, Pogány L, Bakonyi I. Electrochimica Acta 2007;52:3813.

[9] Weihnacht V, Péter L, Tóth J, Pádár J, Kerner Zs, Schneider CM, Bakonyi I. Journal of Electrochem Soc 2003;150:C507.

[10] Péter L, Katona GL, Berényi Z, Vad K, Langer GA, Tóth-Kádár E, Pádár J, Pogány L, Bakonyi I. Electrochimica Acta 2007;53:837.

[11] Bartók A, Csik A, Vad K, Molnár Gy, Tóth-Kádár E and Péter L. submitted to J Phys Chem C, 2009.

**Figure captions**

Figure 1. A crater profile in an electrodeposited multilayer stack created by 350 eV $Ar^+$ ion bombardment.

Figure 2. Intensities of the Cu, Co and Cr signals as a function of depth for both multilayer samples. For the sake of clarity, the elements of the wafer and the support (i.e., Si and Ni, respectively) are not shown, although we detected them during sputtering. The arrows indicate the sputtering direction.

Figure 1. A. Csik, K. Vad, G. A. Langer, G.L. Katona, E. Tóth-Kádár, L. Péter

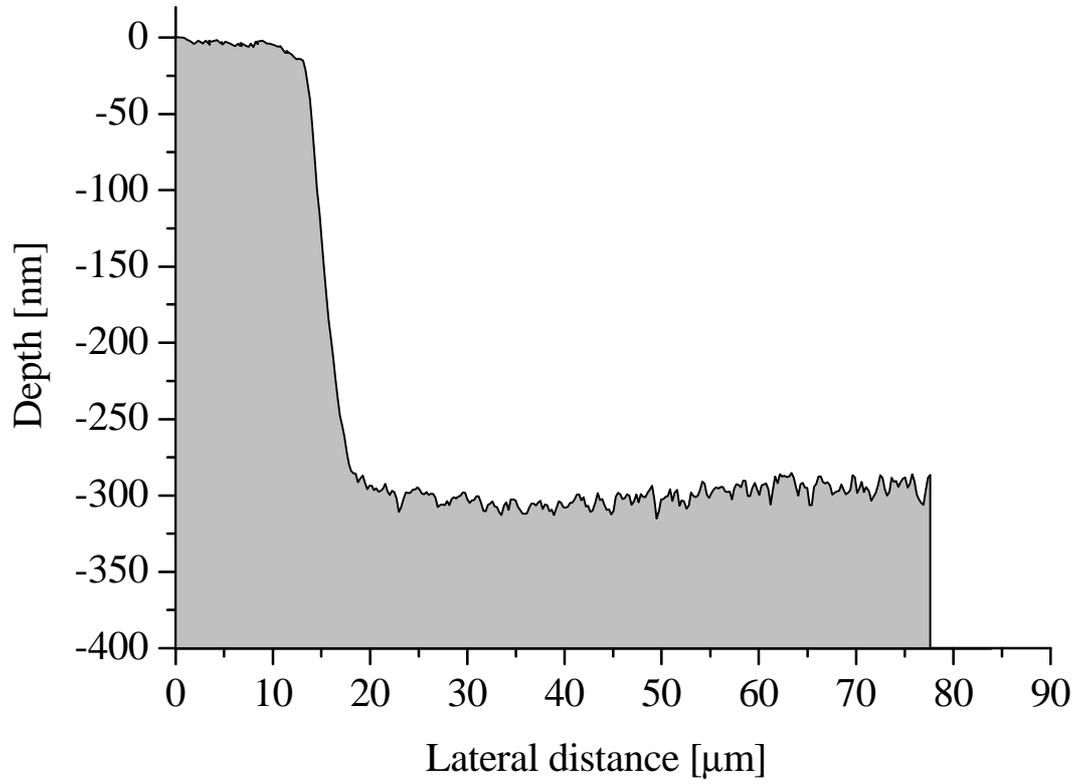

Figure 2. A. Csik, K. Vad, G. A. Langer, G.L. Katona, E. Tóth-Kádár, L. Péter

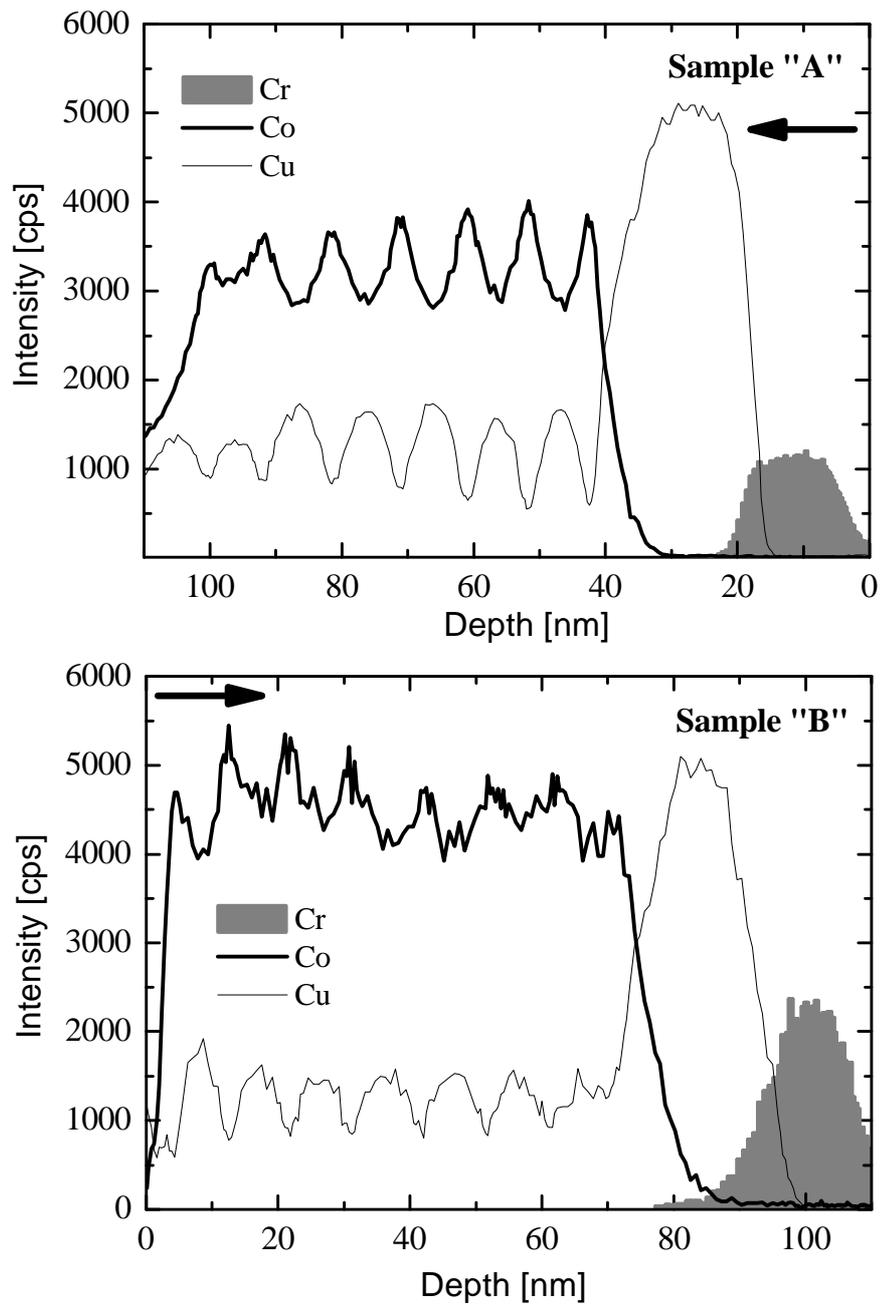